\documentclass[12pt,english,floatfix,superscriptaddress,aps,prd,preprint,nofootinbib]{revtex4}
\usepackage{amsmath}
\usepackage{amssymb}
\usepackage{amsbsy}
\usepackage{amsfonts}
\usepackage{amsopn}
\usepackage{amstext}
\usepackage{graphicx}
\usepackage[english]{babel}
\usepackage{color}
\usepackage{slashed}
\usepackage{esint}
\usepackage[dvips]{epsfig}
\usepackage[dvips]{graphicx}
\usepackage{float}
\usepackage{units}
\usepackage{textcomp}
\usepackage{wasysym}
\usepackage{hyperref}
\usepackage{slashed}
\usepackage{subcaption}
\usepackage{tikz}            
\usetikzlibrary{arrows,quotes,angles,decorations.markings,decorations.pathmorphing} 

\begin{document}

\title{Gravitational lensing by $k-n$ generalized black-bounce space-times}

\author{C. Furtado}
\email{furtado@fisica.ufpb.br}
\affiliation{Departamento de Física, Universidade Federal da Paraíba, Caixa Postal 5008, 58051-970, João Pessoa, Paraíba,  Brazil.}

\author{A. L. A. Moreira}
\email{toinho.alam@gmail.com}
\affiliation{Departamento de Física, Universidade Federal da Paraíba, Caixa Postal 5008, 58051-970, João Pessoa, Paraíba,  Brazil.}

\author{J. R. Nascimento}
\email{jroberto@fisica.ufpb.br}
\affiliation{Departamento de Física, Universidade Federal da Paraíba, Caixa Postal 5008, 58051-970, João Pessoa, Paraíba,  Brazil.}

\author{A. Yu. Petrov}
\email{petrov@fisica.ufpb.br}
\affiliation{Departamento de Física, Universidade Federal da Paraíba, Caixa Postal 5008, 58051-970, João Pessoa, Paraíba,  Brazil.}

\author{P. J. Porfírio}
\email{pporfirio@fisica.ufpb.br}
\affiliation{Departamento de Física, Universidade Federal da Paraíba, Caixa Postal 5008, 58051-970, João Pessoa, Paraíba,  Brazil.} 

\begin{abstract}
 We study gravitational lensing by $k-n$ generalized black-bounce space-times both in regimes of weak and strong field approximations. These metrics interpolate between regular black holes and one-way or traversable wormholes. First, we investigate the light-like geodesic trajectories and derive an analytical expression for the deflection angle in terms of the bounce parameter in the weak-field gravitational regime. We then turn to the strong-field gravitational regime and display the behavior of the bending angle as a function of both the impact parameter and the bounce parameter. Next, using the lens equations, we analyze how the observables for \textit{Sagittarius} A* behave concerning the bounce parameter.  We obtain the shadow's radii for some black-bounce metrics and plot the graph of their sizes, comparing them with the Schwarzschild one.
\end{abstract}

\maketitle

\section{Introduction}

It is undeniable that general relativity (GR) remains our best-tested theory of gravity. This statement is underpinned by its remarkable success in describing numerous gravitational phenomena, such as classical tests and gravitational waves. One notable gravitational phenomenon is gravitational lensing -- the light deflection in the neighborhood of a massive object (often referred to as a gravitational lens). In fact, this effect was originally observed over one hundred years ago \cite{Dyson:1920cwa} (the history of this discovery is presented in various papers, see, for example, \cite{Crispino:2020txj,ademir}). Recent breakthroughs, such as obtaining the image of the shadow of a supermassive black hole at the center of the M87 galaxy \cite{EventHorizonTelescope:2019dse,EventHorizonTelescope:2019uob,EventHorizonTelescope:2019jan,EventHorizonTelescope:2019ths,EventHorizonTelescope:2019pgp,EventHorizonTelescope:2019ggy}, and, furthermore, the observation of the supermassive black hole \textit{Sagittarius} $A^*$ located at the center of our galaxy\cite{Vagnozzi:2022moj}, have further increased interest in the study of gravitational lensing.

{Due to mathematical difficulties, the first works regarding gravitational lensing were carried out in the limit when the light passes far from the gravitational lens ({given by some} massive object) -- this approximation is sometimes referred to as the weak gravitational field regime \cite{Einstein, Liebes, Refsdal}. Nevertheless, the weak-field regime fails to apply when light experiences an intense gravitational field; this is known as the strong gravitational field regime. In {the} latter case, a typical example arises as the light passes near the photon sphere -- that is, a region where the gravitational pull is so strong that the circular lightlike geodesics become unstable -- of a black hole. In this limit, several works have been performed to compute the angular deflection in different contexts, including quantum gravity inspired black holes \cite{Soares:2025hpy, Kumar:2023jgh, Heidari:2024bkm}, non-commutative black holes \cite{Filho:2024zxx, Filho:2024isd, AraujoFilho:2024mvz}, Reissner-Nordstr\"{o}m black holes \cite{Eiroa, Eiroa2, Tsukamoto:2016oca}, wormholes \cite{chetouani, Perlick:2003vg, Nandi:2006ds, Dey:2008kn, Bhattacharya:2010zzb, Nakajima:2012pu, Tsukamoto:2012xs, Nandi:2016ccg, Tsukamoto:2016qro, Tsukamoto:2017edq, Tsukamoto:2016zdu, Shaikh:2018oul, Shaikh:2019jfr, Aounallah:2020rlf}, spinning black holes \cite{Aazami:2011tu, Aazami:2011tw, Bozza:2007gt, Bozza:2005tg, Bozza:2006nm,Bozza:2002af, Chen:2016hil} }. The keystone of the theoretical studies of {strong} gravitational lensing has been put forward by the seminal paper \cite{Virbhadra:1999nm}, where strong gravitational lensing by a Schwarzschild black hole has been studied. The authors found a surprising effect: an infinite sequence of relativistic images, aside from the primary and secondary images, has been generated along the optic axis, which is a purely effect of the strong field regime. Further continuation of this study has been performed in \cite{Claudel:2000yi} where the concept of photon surface generalizing the photon sphere was proposed. Afterwards, several mathematically consistent approaches to the formal study of gravitational lensing in the strong field regime were developed in \cite{Frittelli:1999yf,Bozza:2001xd,Bozza:2002zj, tsukamoto}. Recent results for strong deflection of particles in various contexts are presented in \cite{Feleppa:2024vdk,Feleppa:2024kio,Furtado:2020puz,Tsukamoto:2022uoz, Soares:2024rhp, Soares:2025hpy}, and some other issues related to gravitational lensing can be found also in \cite{Virbhadra:2024xpk,Virbhadra:2022ybp}.

It is well known that GR presents pathological solutions, for example, i) the G\"{o}del solution in which global causality is broken, and ii) solutions that display singularities, {such} as the Schwarzschild black hole. Regarding the latter, Bardeen originally proposed a black hole solution without singularity, the so-called regular black hole \cite{1968qtrconf87B}. Other regular black hole solutions have been obtained in the literature \cite{Bogojevic:1998ma, Hayward:2005gi,Bambi:2013ufa, Toshmatov:2014nya,Azreg-Ainou:2014pra,Ayon-Beato:1998hmi}; in particular, we highlight a class of metrics known as black-bounce ones, which describe non-singular solutions \cite{Simpson:2018tsi, Simpson:2019cer}. Furthermore, regular black holes have the peculiar property of being sourced by a matter content that violates the energy conditions in GR \cite{Fan:2016hvf}. Although this is a stringent constraint in GR, it generically does not hold in alternative theories of gravity scenarios. Within the light deflection context, an important study of regular black holes has been carried out in \cite{nascimento}, the effect of weak and strong gravitational lensing has been studied in the Simpson-Visser (SV) black-bounce space-time, originally introduced in \cite{Simpson:2018tsi}. 
In this paper, we do the next step along this line considering a more generic regular metric called $k-n$ generalized black-bounce space-time \cite{loboetal} which reduces to the one studied in \cite{nascimento} for special values of the parameters. For this metric, we consider the light deflection both for weak-field and strong-field limits. 

The structure of the paper is as follows. In Section \ref{blackbounce}, we briefly describe the main properties of generalized black-bounce space-times. In Section \ref{weak}, we consider the light deflection in the weak field regime {within these regular space-times and then we compute its correspondent deflection angle. In Section \ref{strongfield}, employing Tsukamoto's method, we plot the graphs of angular deflection in terms of the impact parameter. Section \ref{gravitationallensing}} is devoted to a detailed discussion of gravitational lensing {by the generalized black-bounce metrics and to its applications, such as obtaining observables from observational data of \textit{Sagittarius $A^*$} and investigating the sizes of the shadows}.  We summarize our results in Section {\ref{conclusions}}.

\section{Generalized black-bounce space-times}\label{blackbounce}
{One of the most interesting studies of the} light deflection {has been performed} in the standard black-bounce space-time \cite{nascimento,tsukamotoSV}, also known as the SV space-time, {which represents itself as one of the simplest examples of a regular black hole (various issues relating to this solution, including a discussion of corresponding matter sources, can be found in \cite{Bronnikov:2021uta})}. 
 {Later, various generalizations of this metric have been introduced, see, e.g.,} \cite{loboetal}. They are characterized by the typical form of the line element:
\begin{equation}
\label{genmetric}
    ds^2 = f(r) dt^2 - \frac{dr^2}{f(r)} - \Sigma^2(r) (d\theta^2 + \sin^2\theta d\phi^2),
\end{equation}
where $\Sigma(r) = \sqrt{r^2+a^2}$ and $f(r)=1-\frac{2M(r)}{\Sigma(r)}$ \cite{loboetal}. The {dependence on} the {bounce} parameter $a$, {allowing to avoid {the} presence of a singularity,} is the main feature of all black-bounce space-times. It corresponds to a throat at the origin, which may or may not be surrounded by an event horizon. Depending on the value of {$a$, the metric \eqref{genmetric}}  {can} describe a regular black hole with a space-like throat leading into a future copy of the universe, a one-way wormhole if the throat is light-like, or even a traversable wormhole if the throat is time-like. These space-times {present} mostly the same properties as {SV ones}, but some of them {can} have more than one horizon, and some of them {can} violate only some of the energy conditions {required} in GR instead of all of them. It is possible that these metrics do not violate the energy conditions in some modified theories of gravitation. 

The main difference {of the metric (\ref{genmetric}) from the SV one} is the generalization of the {constant} mass to a function of the radial coordinate $M(r)$, which can carry additional parameters and {generate new effects}.
The {example} of space-time {proposed} in \cite{loboetal} introduces the parameters $k$ (a nonnegative integer) and $n$ (a positive integer) by defining
\begin{equation}
\label{massfunction}
    M(r) = \frac{m \Sigma(r) r^k}{(r^{2n} + a^{2n})^{\frac{k+1}{2n}}}.
\end{equation}
The SV {metric} corresponds to the case with $n=1$ and $k=0$.  In any of the new black-bounce space-times, $a\to0$ recovers the Schwarzschild space-time. These metrics have the same restrictions in GR as {the SV one}, i.e., {in general} they do not satisfy the energy conditions, requiring some source of exotic matter whose distribution through space is discussed in detail in \cite{loboetal}, but they have the advantage that this mass function {$M(r)$ can be modified to avoid violating at least some energy conditions, in} a modified theory of gravitation. 

Our goal hereby is to calculate the light deflection and consequently gravitational lensing effects in some of the generalized black-bounce space-times. A more specific characterization of some of these metrics {will be} given in {Section} \ref{strongfield} where we study the photon spheres for the metrics with $k=0, n=2$ and $k=2,n=1$.

\section{Light deflection: weak field regime}\label{weak}
The trajectory of the photon moving through the space-times described by the metric (\ref{genmetric}), with the mass function {given} by (\ref{massfunction}) is determined by the Lagrangian
\begin{equation}\label{lagrangian}
{\cal L}=   \left( 1-\frac{2 m r^k}{(r^{2n} + a^{2n})^{\frac{k+1}{2n}}} \right)\dot{t}^2 - \left( 1-\frac{2 m r^k}{(r^{2n} + a^{2n})^{\frac{k+1}{2n}}} \right)^{-1}\dot{r}^2 - (r^2+a^2) (\dot{\theta}^2 +\sin^2\theta \,\dot{\phi}^2).
\end{equation}
{For the light-like particle we have the constraint ${\cal L}=0$.}
Using the Euler-Lagrange equation for $t$, we {obtain the following expression for the photon energy}:
\begin{equation}\label{energy}
E=\left( 1-\frac{2 m r^k}{(r^{2n} + a^{2n})^{\frac{k+1}{2n}}} \right)\dot{t}.
\end{equation} 
{The evident homogeneity of time allows us to conclude} the energy of the photon is conserved, which {is natural since} the metric is static. {Further}, from the Euler-Lagrange equation {for} $\phi$, we see that the {$z$-component of angular momentum of the photon, given by}
\begin{equation}\label{angularmomentum}
L=(r^2+a^2)\sin^2{\theta}\,\dot{\phi}
\end{equation}
{is also conserved. As a consequence}, the trajectory of the photon is {restricted to} a single {equatorial} plane, {and we can assume $\theta=\frac{\pi}{2}$}. {Taking into account the spherical symmetry of the metric, we can easily find the trajectory of the photon.}

\begin{figure}[htb!]
        \centering
\begin{subfigure}{.45\columnwidth}
    \centering
    \includegraphics[width=.95\linewidth]{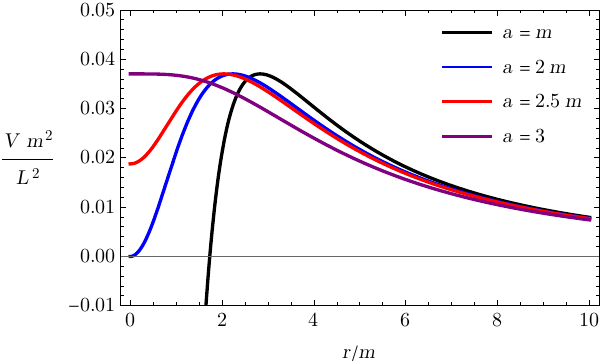}
    \caption{$n=1, k=0$ (SV).}
    \label{potentialsv}
\end{subfigure}
\begin{subfigure}{.45\columnwidth}
    \centering
    \includegraphics[width=.95\linewidth]{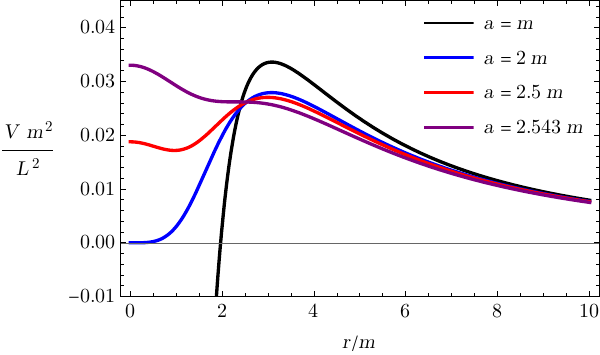}
    \caption{$n=2, k=0$.}
    \label{potentialn2}
\end{subfigure}
\begin{subfigure}{.45\columnwidth}
    \centering
    \includegraphics[width=.95\linewidth]{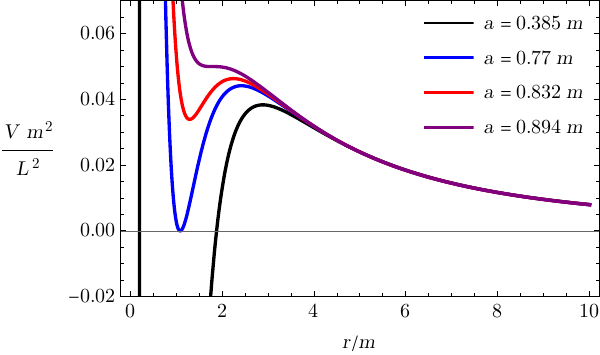}
    \caption{$n=1, k=2$.}
    \label{potentialk2}
\end{subfigure}
\caption{Effective potentials for some specific black-bounce spacetimes.}
\label{potentials}
\end{figure}

If we substitute \eqref{energy} and \eqref{angularmomentum} in \eqref{lagrangian}, and {require $\theta=\pi/2$}, we get
\begin{equation}\label{trajectoryequation1}
    E^2 = \dot{r}^2 + L^2 \left( 1-\frac{2 m r^k}{(r^{2n} + a^{2n})^{\frac{k+1}{2n}}} \right)(r^2+a^2)^{-1}
\end{equation}
{So, the problem of finding the trajectory of light reduces to that of a one-dimensional movement for a particle under the effective potential $V(r) \equiv L^2 \frac{f(r)}{\Sigma^2(r)}$} {displayed at Fig. \ref{potentials}}.

Each photon coming from infinity has an impact parameter $b=\frac{L}{E}$, which determines its behavior in response to $V$, and consequently its trajectory. Depending on its value, the photon {can either} be scattered {in some direction}, fall into the black hole, or, in the strong field limit, stay orbiting {around it} indefinitely in the photon sphere. There is a {maximal} approximation point ($r_0$) {that can be found for} every impact parameter. In other words, when a photon comes from infinity, it passes near the black hole at a {minimal} distance $r_0$ and is scattered in another direction afterward.
The angle between its {trajectories before and after scattering} is the deflection angle {which we will denote as} $\alpha$.

\begin{figure}[htb!]
        \centering
\begin{subfigure}{.45\columnwidth}
    \centering
    \includegraphics[width=.95\linewidth]{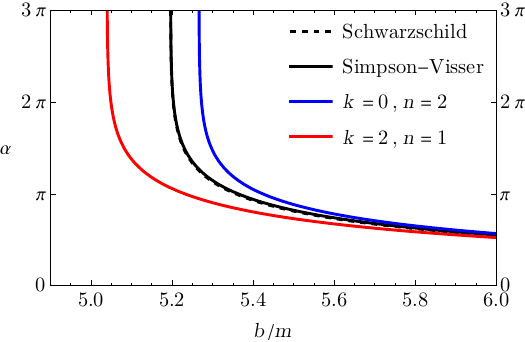}
    \caption{$a=\frac{1}{2}m$}
    \label{defla.5}
\end{subfigure}
\begin{subfigure}{.45\columnwidth}
    \centering
    \includegraphics[width=.95\linewidth]{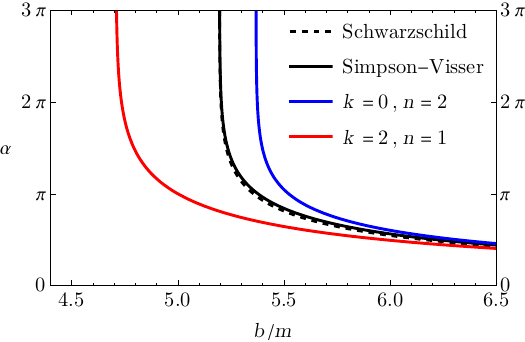}
    \caption{$a=\frac{4}{5}m$}
    \label{defla1}
\end{subfigure}
\begin{subfigure}{.45\columnwidth}
    \centering
    \includegraphics[width=.95\linewidth]{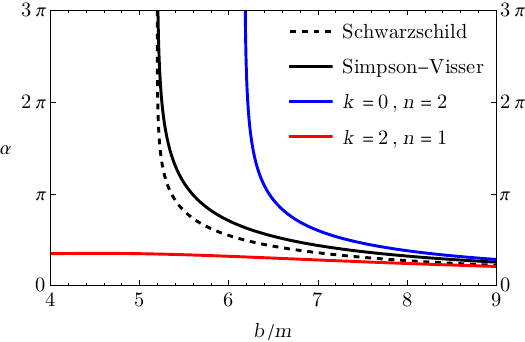}
    \caption{$a=\frac{5}{2}m$}
    \label{defla2.5}
\end{subfigure}
\caption{Light deflection as a function of the impact parameter for different space-times with the same value of $a$.}
\label{defla}
\end{figure}

Since the impact parameter is constant for the entire trajectory, its value is easily obtained by {knowing} $r_0$, because at this point $\dot{r}=0$, and from \eqref{trajectoryequation1} we have $b^2=\frac{\Sigma^2(r_0)}{f(r_0)}$.
Rewriting \eqref{trajectoryequation1} with {use of} the definitions of the effective potential $V$, the impact parameter $b$ and the angular momentum $L$, we obtain a trajectory equation relating $r$ and the angular coordinate $\phi$:
\begin{equation}
    \dot{r}^2 = \dot{\phi}^2\, \Sigma^4(r) \left(\frac{1}{b^2}-\frac{f(r)}{\Sigma^2(r)}\right)
\end{equation}
We can then define $\alpha$ from this equation by integrating the angle {where radius varies} from $r_0$ to $\infty$ to obtain the first branch of the trajectory, {multiplying the result by 2,} and then subtracting $\pi$, {so, finally we obtain}
\begin{equation}\label{alpha}
    \alpha=2\int_{r_0}^\infty \frac{dr}{\sqrt{\frac{\Sigma^4(r)\,f(r_0)}{\Sigma(r_0)}-f(r)\Sigma^2(r)}} - \pi.
\end{equation}
This {integral cannot be expressed in closed form}, so, it {must} be approximated in some way. {It can be found} using some numerical approximation, as in {Fig.} \ref{defla}.
{At the same time, several physically interesting results can be obtained through expansions of this integral.} We first consider light deflection in the weak field limit, that is, when $m << 1$, and consequently $a<<1$. This {scenario corresponds to the large distance} from the center of the black hole. {In} the zeroth order, {we have zero result which is natural} since the black hole should not affect regions {very} distant from it. In this limit $r_0\approx b$, so we express the result in terms of $b$.
{Then, we can write down the Taylor series up to the} second order, {to obtain} the deflection in every black-bounce space-time:
\begin{equation}
\alpha \approx \frac{4m}{b}+\frac{(15\pi-16)m^2}{4b^2}+\frac{\pi a^2}{4b^2}
\end{equation} 
as is the case for SV {space-time}. Therefore we conclude that in the weak field limit the new black-bounce space-times cannot be {distinguished} from the SV {space-time}.

\section{Light deflection: Strong field regime}\label{strongfield}

Now we {turn to} the strong field limit. {In this case, we give main attention to the light deflection} near the photon sphere, whose {radius} is $r_m$. Since these metrics {are spherically symmetric, static {and asymptotically flat}}, we can use the same procedure {developed} in \cite{tsukamoto} to calculate the {deflection of light in our case}. The first step is to calculate the radius of the photon sphere ($r_m$), for which the following expression is valid:
\begin{equation}\label{eqphotonspheres}
    V'(r_m) = 0.
\end{equation}
In this expression, the prime denotes the derivative with respect to the coordinate $r$.
This expression {itself} could {describe} a maximum or a minimum of the effective potential, but for the metrics studied {here,} the outermost photon sphere, or the {largest} solution of \eqref{eqphotonspheres} is always a maximum. 
{Indeed,} if {there was} a minimum, the photon would have a greater energy than is needed to be caught in the potential well and would pass right through it.
Where there is a maximum potential, it becomes an unstable equilibrium point. {If} the innermost radius attained by the photon ($r_0$) coincides with the photon sphere ($r_m$), the photon becomes trapped in it forever {moving} around the black hole. In practice, two situations occur, {namely: if $r_0<r_m$, the photon follows an {unstable circular} orbit and then falls inside the black hole, and if $r_0>r_m$, the photon is scattered by the black hole to infinity.} {{As} its energy approaches its maximum, it circles the black hole an increasing number of times}. This {scenario} will become more important in Section \ref{gravitationallensing}. Now we are concerned with the {dependence} of the deflection angle on the parameters of the metric.

\begin{figure}[ht]
        \centering
        \label{photonspheres}
\begin{subfigure}{.45\columnwidth}
    \centering
    \includegraphics[width=.95\linewidth]{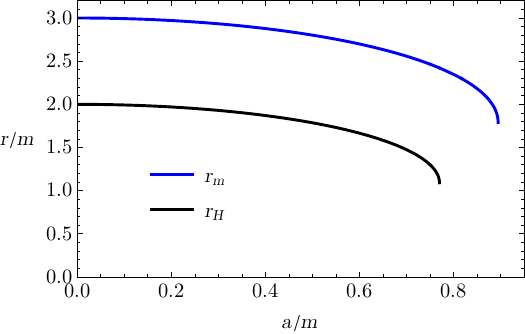}
    \caption{$k=2$, $n=1$}
    \label{photonspheresk2n1}
\end{subfigure}
\begin{subfigure}{.45\columnwidth}
    \centering
    \includegraphics[width=.95\linewidth]{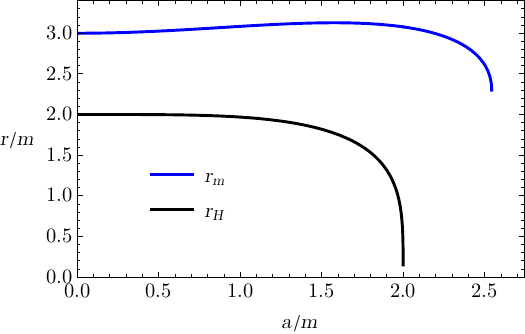}
    \caption{$k=0$, $n=2$}
    \label{photonspheresk0n2}
\end{subfigure}
\caption{Photon sphere (blue) and event horizon (black) for black-bounce space-times. }
\end{figure}

{To start, let us} consider the case with $k=2$ and $n=1$ (Fig. \ref{photonspheresk2n1}) and the case with $k=0$ and $n=2$ (Fig. \ref{photonspheresk0n2}).
Other space-times with $n=1$ and $k\neq 0,2$ have configurations of event horizon and photon sphere {analogous} to \ref{photonspheresk2n1}. {Similarly}, other space-times with $n>1$ and for all $k$ behave like \ref{photonspheresk0n2}. 
  The differences among the different kinds of black-bounce metrics are more evident as we approach the photon spheres.

{Let us now proceed with an in-depth assessment of the horizon and photon sphere structures of some particular black-bounce metrics. To gain more insight on this, we first consider the} SV {metric}, {which possesses the following features:
\begin{itemize}
    \item a horizon at $r_H=\sqrt{(2m)^2-a^2}$ and a photon sphere at $r_m=\sqrt{9m^2-a^2}$ for $0\leq a<2m$;
    \item a marginal horizon at $r_H=0$ and a photon sphere at $r_m=\sqrt{5}m$ for $a=2m$;
    \item no horizon and a photon sphere for $2m<a<3m$;
    \item no horizon and a marginal photon sphere at $r_m=0$ for $a=3m$;
    \item no horizon and no photon sphere for $a>3m$.
\end{itemize}

}
 {Now,} let us explore two different cases for the generalized black-bounce metrics, namely: 
\begin{itemize}
\item for $n=2$ and $k=0$ and $a<2m$, we found one horizon at $r_H=\sqrt[4]{(2m)^4-a^4}$ and one photon sphere whose radius is given by the equation
\begin{equation}
    3mr_m^5+ma^2r_m^3+2ma^4r_m=(r_m^5+a^4r_m)\sqrt[4]{r_m^4+a^4},
\end{equation} for $a=2m$ it has a {marginal} horizon at $r_H=0$ besides one photon sphere, for $2m<a<2.54343m$ it has an inner and an outer photon sphere, for $a=2.54343m$ there is only one photon sphere at $r_m=2.29825m$ and for $a>2.54343m$ there are no photon spheres and no horizons.
\item for $n=1$ and $k=2$, one has two horizons for $a<\frac{4m}{3\sqrt3}$ and two photon spheres for $a<\frac{2m}{\sqrt5}$ while for $a=\frac{2m}{\sqrt5}$ it has one photon sphere at $r_m=\frac{4m}{\sqrt{5}}$ and for $a>\frac{2m}{\sqrt5}$ it has no photon spheres and no horizons.
\end{itemize}

If we simply {substitute} $r_m$ {for} $r_0$ {into} \eqref{alpha}, the integral diverges \cite{tsukamoto}, which is expected because {a photon} becomes trapped in the photon surface. 
{Thus, the light deflection} angle $\alpha$ is calculated with $r_0 \to r_m$ in the strong field limit.
When $r_0 \to r_m$, the impact parameter approaches a critical value ($b_c$), so we can also express the deflection angle in terms of $b_c$ instead of $r_m$, which will become more convenient in Section \ref{gravitationallensing} where we calculate the observables. {We can compare graphically how $b_c$ changes with $a$ among the black-bounce spacetimes (see Fig. \ref{plotbc}):
\begin{figure}[htb!]
    \centering
    \includegraphics[width=.5\linewidth]{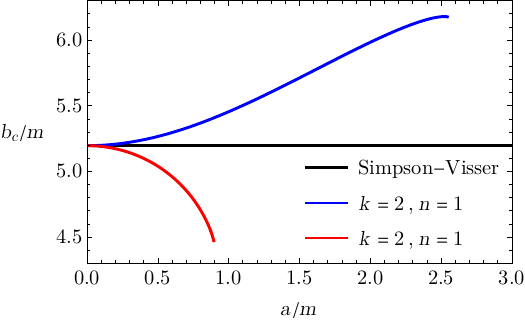}
     \caption{Critical impact parameter as a function of $a$.}
    \label{plotbc}
\end{figure}
The only black-bounce metric with a closed form {of the} critical impact parameter is the SV metric, for which $b_c=3\sqrt{3}m$, the same as Schwarzschild, independent of $a$. for the other two black-bounce metrics, $b_c$ depends on the value of $a$.}

\begin{figure}[htb!]
    \centering
    \includegraphics[width=.5\linewidth]{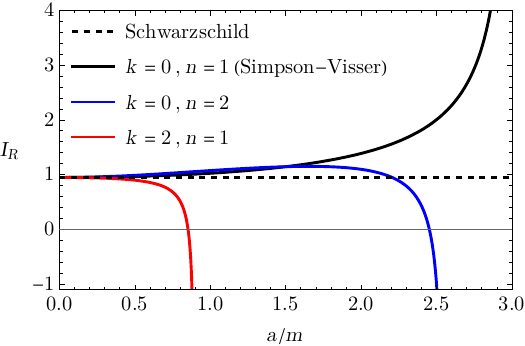}
     \caption{Regular integral as a function of parameter $a$.}
      \label{plotintegralregular}
\end{figure}

Since the integral diverges in this case, the method devised in \cite{tsukamoto} can be used to extract the divergent part of the {result}.
{It remains to evaluate} the regular part of the integral. The result {looks like}:
\begin{equation}\label{desviodaluz}
    \alpha(b) \approx -\mathcal{A} \ln{\left(\frac{b}{b_c}-1\right)} + \mathcal{B},
\end{equation}
{where} the coefficients $\mathcal{A}$ and $\mathcal{B}$ are constants, and the logarithmic term diverges in the strong field limit ($b \to b_c$). The constant $\mathcal{A}$ is given by:
\begin{equation}
    \mathcal{A} = \frac{1}{\sqrt{f_m-\frac12 f''_m \Sigma_m^2}}.
\end{equation}
And the constant $\mathcal{B}$ {looks like}:
\begin{equation}
        \mathcal{B} = \mathcal{A} \ln{\bigg(r_m^2\Big(\frac{2}{\Sigma_m^2}-\frac{f''_m}{f_m}\Big)\bigg)} + I_R - \pi.
\end{equation}
In this expression, $I_R$ is the regular part \eqref{parteregular} of the integral of the light deflection (\textit{cf.} \eqref{alpha}):
\begin{equation}
        \label{parteregular}
    I_R=\int_0^1{dz\left(\frac{2 r_m}{(1-z)^2 \sqrt{\frac{f_m}{\Sigma_m^2}\Sigma^4(\frac{r_m}{1-z})-\Sigma^2(\frac{r_m}{1-z}) f (\frac{r_m}{1-z})}}-\frac{2}{z \sqrt{f_m-\frac12 f_m '' \, \Sigma_m^2}}\right)}
\end{equation}
where $z=1-\frac{r_0}{r}$ as defined in \cite{tsukamoto}. 
The first term inside the integral is {just} the integrand of \eqref{alpha} after substituting $z$ for $r$ and $r_m$ for $r_0$. 
The second term inside the integral is the divergent term, which is used to calculate the strong field coefficients $\mathcal{A}$ and $\mathcal{B}$, so we subtract the divergent term from the original integral to obtain a finite result and the divergent part becomes a logarithmic function which tends to infinity. For the metric with $k=0,n=2$, this expression becomes
\begin{multline}
    \int_0^1\frac{2r_m dz}{\sqrt{\big(r_m^2+a^2(z-1)^2\big)\Bigg(\frac{\big(r_m^2+a^2(z-1)^2\big)\big(\sqrt[4]{r_m^4+a^4}-2m\big)}{(r_m^2+a^2)\sqrt[4]{r_m^4+a^4}}-\frac{2m(z-1)^3}{\sqrt[4]{r_m^4+a^4(z-1)^4}}-(z-1)^2\Bigg)}}\\-\int_0^1\frac{2(r_m^4+a^4)^{\frac98} dz}{z\sqrt{2a^2m r_m^6-7a^4m r_m^4-3a^6m r_m^2-2a^8m+(r_m^4+a^4)^{\frac94}}},
\end{multline}
{where, as a result of Eq.\eqref{eqphotonspheres}, $r_m$ is given by the {following} equation
\begin{equation}
    \left(a^4+r_m^4\right)^{5/4}-m \left(2 a^4+a^2 r_m^2+3 r_m^4\right)=0.
\end{equation}
} 
And for $k=2,n=1$ the regular integral is written as
\begin{multline}
    \int_0^1 \frac {2r_m dz} {
    \sqrt{
    \frac{
    \big( r_m^2 + a^2 (z-1)^2\big) \big( (r_m^2+a^2)^{\frac32} - 2m r_m^2\big)
    } {(r_m^2 + a^2)^{\frac52} } - 
    r_m^2(z-1)^2 \Big( 1 + \frac{2m(z-1)}{\sqrt{r_m^2 + a^2(z-1)^2 } } \Big) - a^2(z-1)^4)
    }}\\-\int_0^1\frac{2(r_m^2+a^2)^{\frac54}dz}{z\sqrt{ma^2(2a^2-13rm^2)+(r_m^2+a^2)^{\frac52}}},
\end{multline}
{where, as a result of Eq.\eqref{eqphotonspheres}, $r_m$ satisfies the following {equation}
\begin{equation}
    a^2 \left(\sqrt{a^2+r_m^2}+2 m\right)+r_m^2 \left(\sqrt{a^2+r_m^2}-3 m\right)=0.
\end{equation}}
In Fig. \ref{plotintegralregular} we compare the results of the regular integral for the metrics whose photon spheres were previously illustrated with the regular integral for the SV metric.
We can see in this {figure} that as $a$ approaches a limit for which there can no longer be photon spheres, the regular integral diverges, {tending to $+\infty$} for SV space-time and {to $-\infty$} for {the} generalized black-bounce space-times.  For small values of $a$, the {regular integral} does not depend on $n$ and $k$.

\begin{figure}[htb!]
        \centering
\begin{subfigure}{.45\columnwidth}
    \centering
    \includegraphics[width=.95\linewidth]{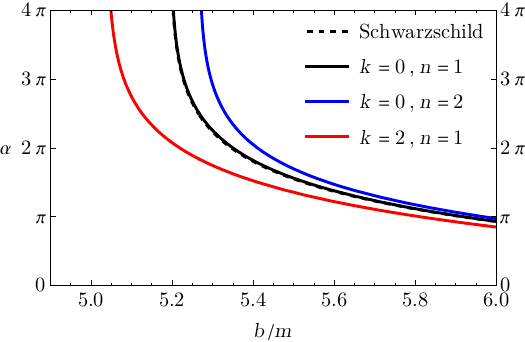}
    \caption{$a=\frac{1}{2}m$}
\end{subfigure}
\begin{subfigure}{.45\columnwidth}
    \centering
    \includegraphics[width=.95\linewidth]{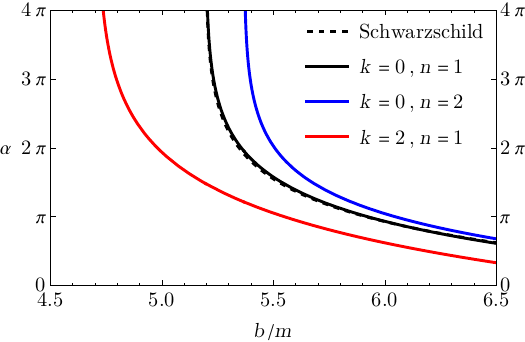}
    \caption{$a=\frac{4}{5}m$}
\end{subfigure}
\begin{subfigure}{.45\columnwidth}
    \centering
    \includegraphics[width=.95\linewidth]{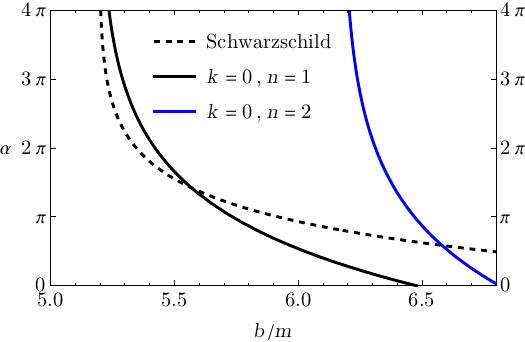}
    \caption{$a=\frac{5}{2}m$}
\end{subfigure}
\caption{Strong field approximation for different space-times and fixed $a$.}
\label{sfdeflection}
\end{figure}

 As {illustrated in Fig. \ref{sfdeflection}, when} we increase $a$, not only the curves {become} different, but the critical impact parameters {and} the value for which the deflection {angle blows up}, {present differences}. {As we restricted ourselves here} to the strong field limit, we did not calculate the deflection for values of $a$ for which the metric does not exhibit photon spheres. This is the reason why there is no $k=2$ curve for $a=\frac{5}{2}m$ (its limit for photon spheres is $a=\frac{2m}{\sqrt{5}}$).

\section{Gravitational lensing}\label{gravitationallensing}

{After we have written} light deflection in the form of \eqref{desviodaluz}, it becomes possible to calculate two physical quantities related to the nearest surroundings of a black hole outside its shadow.
In the strong field limit, relativistic images {are {located very close to}} the main image. {They are formed by photons} from some source behind the black hole whose impact parameter results in a deflection of more than $2\pi$, i.e. they do more than one turn around the black hole before coming out to the observer. Relativistic images are also fainter and more compact than the main image.
There is a relativistic image for each number ($N$) of complete rotations. 
We follow the same procedure adopted in \cite{nascimento,tsukamoto,bozza2002, bozza2001}.
\subsection{Lens equation}
\begin{figure}
\centering
\begin{tikzpicture}[scale=0.9]
\node (I)    at ( 5,-0.4)   {$L$};
\node (I)    at ( 6.7,0.9)   {$\theta_N$};
\node (II)    at ( 10,-0.5)   {$O$};
\node (II)    at ( -0.5,1.5)   {$S$};
\node (II)    at ( -0.5,5)   {$I$};
\node    at ( 2.5,-1.5)   {$D_{LS}$};
\node    at ( 7.5,-1.5)   {$D_{OL}$};
\draw (10,0)--(0,0)--(0,5);
\draw [thick,rounded corners=20pt] (0,1.5)--(5,2.5)--(10,0);
\draw [dashed] (5,2.5)--(0,5);
\draw [dashed](10,0)--(0,1.5);
\fill[orange] (5,0) circle (2pt);
\fill[purple] (10,0) circle (2pt);
\fill[red] (0,1.5) circle (2pt);
\fill[blue] (0,5) circle (2pt);
\draw
(6,2) coordinate (a) 
-- (5,2.5) coordinate (b) 
-- (6,2.7) coordinate (c) 
pic["$a$", draw=blue, <->, angle eccentricity=1.2, angle radius=0.9cm]
{angle=a--b--c};
\draw
(4,2.3) coordinate (d) 
-- (5,2.5) coordinate (e) 
-- (4,3) coordinate (f) 
pic["$a$", draw=blue, <->, angle eccentricity=1.2, angle radius=0.9cm]
{angle=f--e--d};
\draw
(8,0) coordinate (g) 
-- (10,0) coordinate (h) 
-- (8,1) coordinate (i) 
pic[ draw=black, angle eccentricity=1.05, <->, angle radius=2.8cm]
{angle=i--h--g};
\draw
(7,0) coordinate (g1) 
-- (10,0) coordinate (h1) 
-- (7,0.47) coordinate (i1) 
pic["$\beta$", draw=black, angle eccentricity=1.1, <->, angle radius=2.35cm]
{angle=i1--h1--g1};	
\draw[<->, decoration = {markings, mark = at position 0.5 with {\arrow{|} }, }, postaction = {decorate}] (0,-1) node[ anchor = west ] {} -- (10,-1);
\end{tikzpicture}
\caption{Schematic representation of the observer (O), the source (S), the lens (L) and the image (I).}
\label{figlente}
\end{figure}
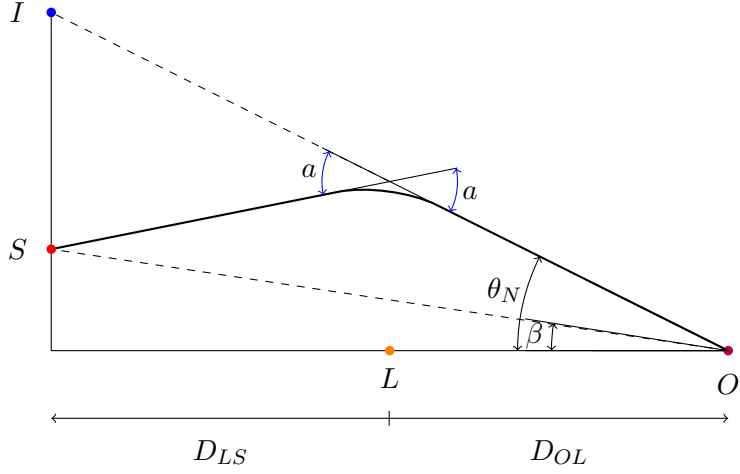

Let us exclude all full rotations by defining $\alpha_N \equiv \alpha - 2\pi N$. Then we can calculate the difference between the {angular positions of} the source ($\beta$) and {of the image seen by observers} ($\theta_N$), illustrated in Fig. \ref{figlente}.
The relation between {these} angles is the Lens Equation \eqref{eqdalente}.
\begin{equation}\label{eqdalente}
       \beta = \theta_N - \frac{D_{LS}}{D_{OS}} \alpha_N.
\end{equation}
{Here} we suppose $D_{OS}\approx D_{LS}+D_{OL}$, for the angle $\alpha_N$ must be tiny {to guarantee the visibility for lensing effects}. In this regard, {it follows from geometrical reasons that} $b\approx D_{OL} \tan{\theta_N} \approx D_{OL} \theta_N$. 

Thus, $\alpha_N$ can be {treated} as a function of $\theta_N$. Then, we expand $\beta$ to first order in $(\theta_N - \theta_N^0)$, with {$\theta_N^0$ defined} as an angle such that $\alpha_N(\theta_N^0) = 0$.
\begin{equation}\label{anguloexp}
    \beta \approx \theta_N^0 +\bigg(1+\frac{\mathcal{A}\, D_{LS} D_{OL}}{b_c\, D_{OS}} 
    e^{
    \frac{2\pi N - \mathcal{B}}{\mathcal{A}}
    } \bigg)(\theta_N - \theta_N^0)
\end{equation}
This expression can also be easily inverted in order to obtain the angular position of the image ($\theta_N$) as a function of the angular position of the source ($\beta$), by noting that in {(\ref{anguloexp}) one can neglect 1 in comparison with another term}. So, {one finds}
\begin{equation}\label{anguloexpinv}
    \theta_N \approx \theta_N^0 + \frac{b_c \, D_{OS}}{\mathcal{A} \, D_{OL} D_{LS}} e^{
    \frac{\mathcal{B} - 2\pi N}{\mathcal{A}}
    }(\beta - \theta_N^0).
\end{equation} 
From \eqref{anguloexp} and \eqref{anguloexpinv} we can define the observables which depend on the deflection coefficients $\mathcal{A}$ and $\mathcal{B}$. 

\subsection{Observables}

The first observable is the separation angle ($s$) between the first ($N=1$) and the remaining relativistic images, all {merged} in a single image denoted by an infinity index ($N \to \infty$). Since the correction in \eqref{anguloexpinv} is much smaller than $\theta_N^0$, $s$ can be approximated as follows:
\begin{equation}
    s \approx \theta_1^0 - \theta_{\infty}^0 = \frac{b_c}{D_{OL}} e^{\frac{\mathcal{B}-2\pi}{\mathcal{A}}}.
\end{equation}
The other observable defined from the deflection coefficients is the ratio between the luminous flux of the first relativistic image and the remaining ones. It depends on the magnification $\mu_N \equiv \left(\frac{\beta}{\theta_N}\frac{\partial \beta}{\partial \theta_N}\right)^{-1}\bigr|_{\theta_N^0}$. Because of the large distances involved the magnification can be approximated as
\begin{equation}
    \mu_N = \frac{b_c\!^2 D_{OS} \,
    e^{\frac{\mathcal{B} - 2\pi N}{\mathcal{A}}}
    \Bigl(1+
    e^{\frac{\mathcal{B} - 2\pi N}{\mathcal{A}} }
    \Bigr)}
    {\beta \, \mathcal{A} \, D_{LS} \, D_{OL}\!^2}.
\end{equation}
Hence, we calculate $R$, and, from the order of magnitude of the quantities considered, {present its approximate value as}:
\begin{equation}
    R \equiv \frac{\mu_1}{\sum\limits_2^{\infty} \mu_N}\\ = 
    \left(e^{\frac{2\pi}{\mathcal{A}}}
    +e^{\frac{\mathcal{A}}{\mathcal{B}}}\right)
    \left(
    \frac{
    e^{\frac{2\pi}{\mathcal{A}}}
    }{
    e^{\frac{2\pi}{\mathcal{A}}}-1
    }
    +\frac{
    e^{\frac{\mathcal{A}}{\mathcal{B}}}
    }{
    e^{\frac{4\pi}{\mathcal{A}}}-1
    }
    \right)
    \approx e^{\frac{2\pi}{\mathcal{A}}}.
\end{equation}
\subsection{Application for \textit{Sagittarius} A*}
\begin{figure}[htb!]
\centering
\begin{subfigure}{.45\textwidth}
    \centering
    \includegraphics[width=1\linewidth]{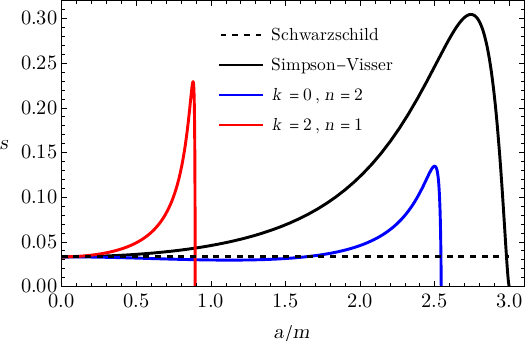}
    \caption{Separation angle in {\textmu}arcsec.}
    \label{figseparacao}
\end{subfigure}
\begin{subfigure}{.45\textwidth}
    \centering
    \includegraphics[width=1\linewidth]{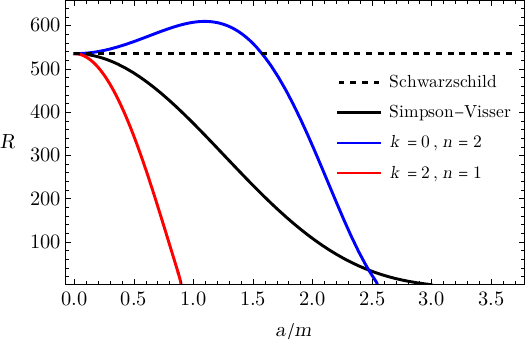}
    \caption{Ratio between luminous fluxes.}
    \label{figratios}
\end{subfigure}
\caption{Observables as a function of parameter $a$ with \textit{Sagittarius} A* acting as a gravitational lens.}
\label{figsepratio}
\end{figure}

In order to compare the results of each metric, we suppose the lens is the black hole at the center of the galaxy, \textit{Sagittarius} A*, with mass ($4.297\times10^6 M_\odot$) and distance (8277 pc) profile taken from \cite{dadosdesagittariusa}. Fig. \ref{figsepratio} presents the separation angles and ratios between luminosities for the metrics studied.
We can see in the graphs that the observables can {differ essentially} from the Schwarzschild case, {if the values of $a$ are} not too small. 


{\subsection{Shadows}

Black hole and wormhole shadows are {typical} gravitational lensing effects that arise in the strong field regime. As discussed in the previous section,  the photon sphere -- with critical radius $r_m$ -- is the region where photons are forced to follow unstable circular orbits. As a result, such orbits cast a dark region on the observer's sky, called the shadow. In particular, for a static and spherically symmetric {metric}, the radius of the shadow regarding an observer at infinity can be expressed as 
\begin{equation}
    r_{s}=\frac{r_m}{\sqrt{f(r_m)}}.
\end{equation}
For the special case of the SV metric, which corresponds to
\begin{equation}
r_m=3\sqrt{m^2-\left(\frac{a}{3}\right)^2}\,\,\,\,\, \mbox{and} \,\, \,\,\,f(r_m)=\frac{1}{3},
\end{equation}
one can find an analytical expression for the radius of the shadow given by
\begin{equation}
    r_s=3\sqrt{3}\sqrt{m^2-\left(\frac{a}{3}\right)^2},
\end{equation}
which recovers the standard result (Schwarzschild black hole) for $a=0$. From the previous equation, it is straightforward that {while} $a$ grows, $r_s$ decreases -- with an upper bound on the bounce parameter given by $a=3m$. Note, however, that depending on the value of $a$,  the spacetime can represent either a black hole or a wormhole. As discussed in Sec. \ref{strongfield}, black holes are recovered for $0<a\leq 2m$, and wormholes with nontrivial photon spheres are obtained for $2m<a\leq 3m$.  

To further illustrate these cases, we explore how the size of the shadows is affected by the bounce parameter $a$. In Fig. \ref {fig:enter-label}, we plot shadow size for the SV metric for different values of $a$, assuming $m=1$. The reference (dashed) line represents the Schwarzschild photon sphere, while the disk $a=0$ describes the size of the Schwarzschild black hole shadow. The other curves, $a=1.5$ and $a=2.5$, describe an SV black hole and an SV wormhole, respectively.
\begin{figure}[htb!]
        \centering
        \includegraphics[width=0.6\linewidth]{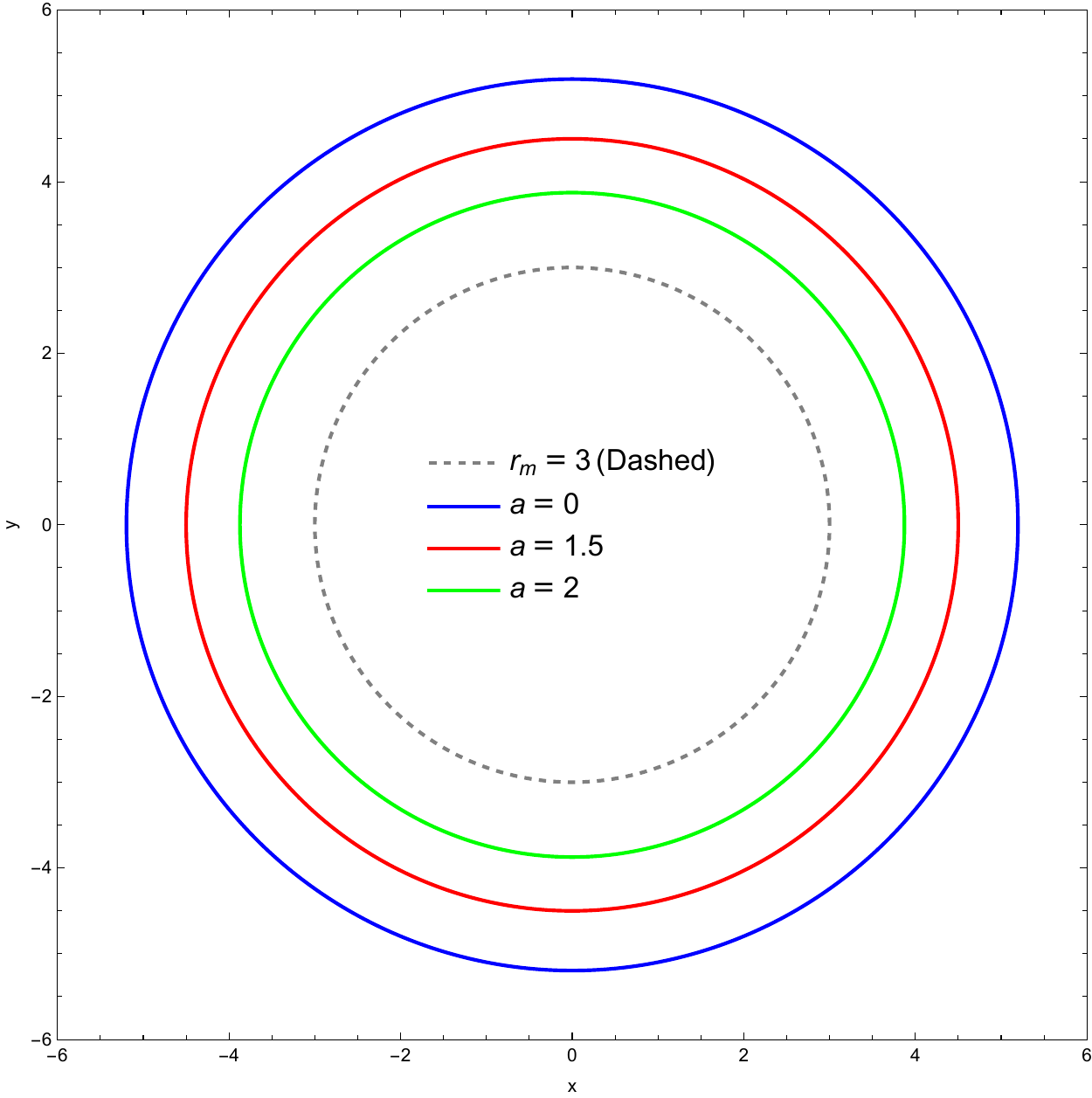}
            \caption{The circles represent the shadow disks for the SV metric for different values of $a$ for fixed $m=1$. In particular, the outermost circle, blue line ($a=0$), is the Schwarzschild shadow disk. The dashed circle stands for the Schwarzschild photon sphere. }
            \label{fig:enter-label}
 \end{figure}

Now, we turn our attention to generalized black-bounce metrics. We illustrate the size of the shadows for $k=2$ and $n=1$, and $k=0$ and $n=2$ generalized black-bounce metrics. For the former case ($k=2$ and $n=1$), Fig.\ref{k2} exhibits a global behavior similar to that of the SV metric. Schwarzschild black hole corresponds to the largest shadow disk (blue line in Fig. \ref{k2}). Notably, the curves, $a=0.5$ and $a=0.7$, describe the shadow disks of black holes, while $a=0.8$ represents a shadow disk of a wormhole.
\begin{figure}[htb!]
        \centering
        \includegraphics[width=0.6\linewidth]{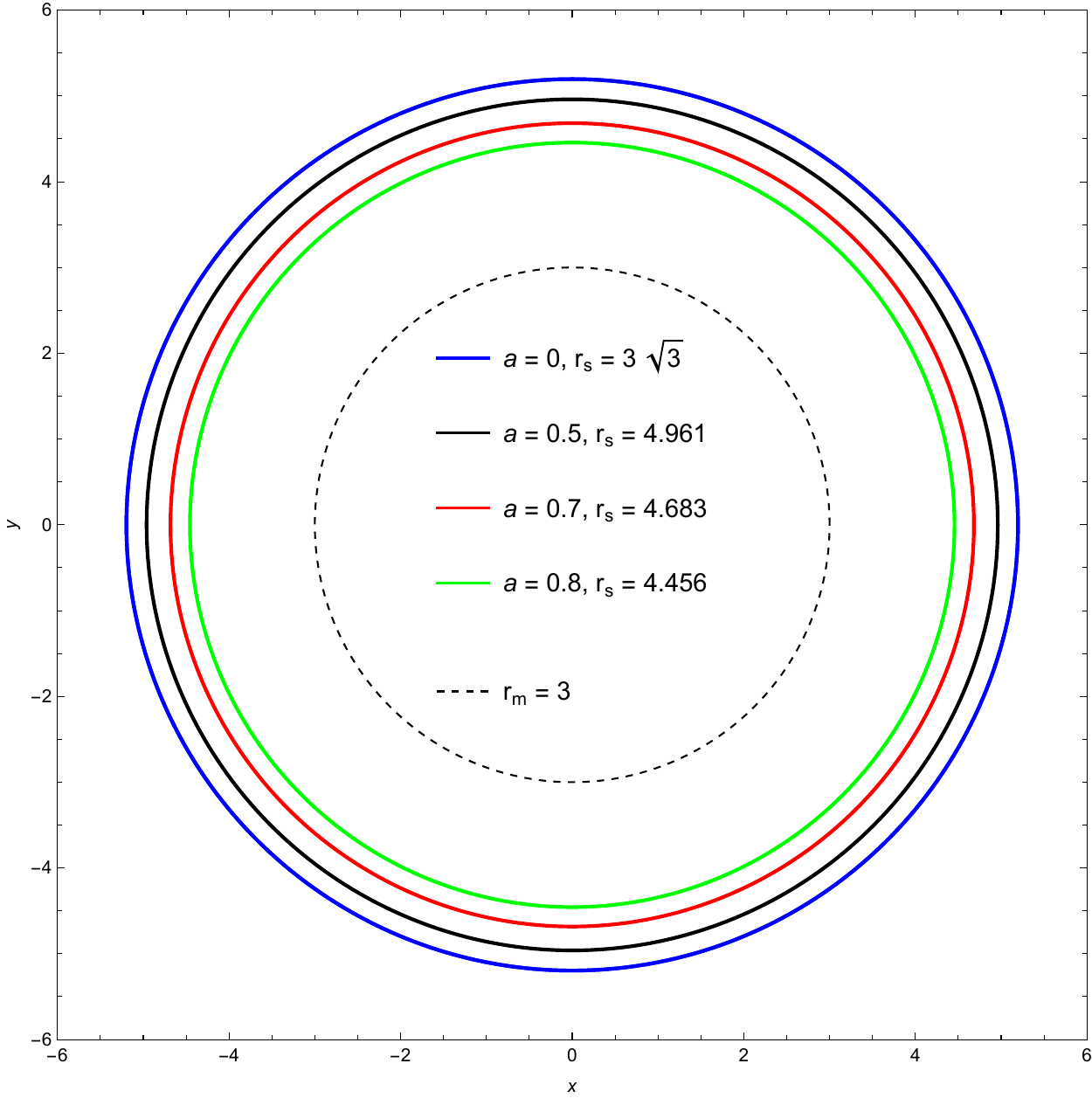}
            \caption{The circles represent the shadow disks for the generalized black-bounce metric ($k=2$ and $n=1$) for different values of $a$ for fixed $m=1$. Similar to the previous case, the outermost circle, blue line ($a=0$), represents the Schwarzschild shadow disk, while the dashed one stands for the Schwarzschild photon sphere. }
            \label{k2}
 \end{figure}  
Fig. \ref{k0} displays the behavior of the shadow disks for the case, $k=0$ and $n=2$. The qualitative results are similar to those of the two previous cases, as can be seen in Fig.\ref{k0}. Again, the outermost shadow disk corresponds to the Schwarzschild black hole. Note that the curves, $a=1$ and $a=1.57532$, correspond to black holes, whereas the curve $a=2.3$ describes a wormhole.

\begin{figure}[htb!]
        \centering
        \includegraphics[width=0.6\linewidth]{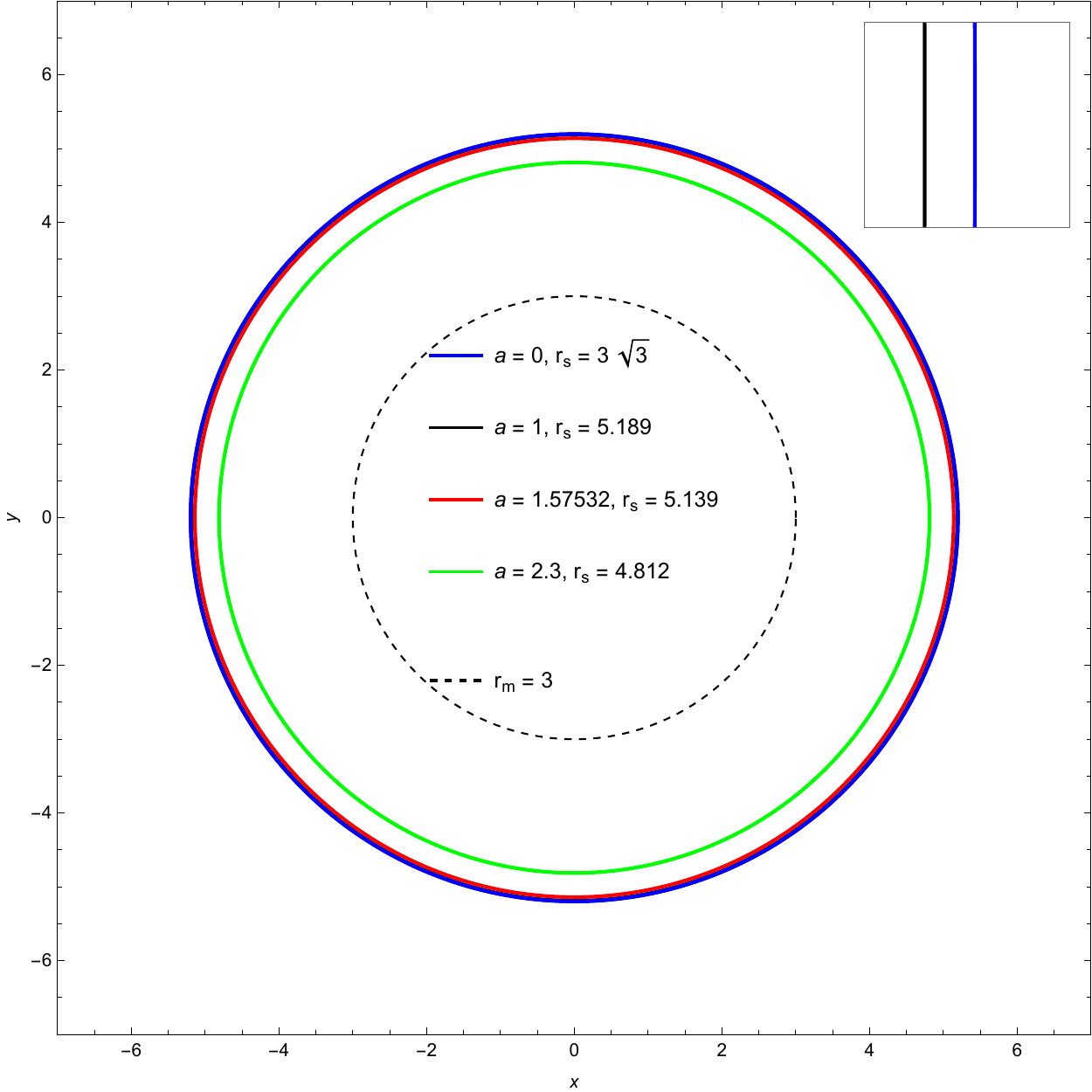}
            \caption{The circles represent the shadow disks for the generalized black-bounce metric ($k=0$ and $n=2$) for different values of $a$ for fixed $m=1$. The inset figure shows the narrow area between the outermost (blue) circle (Schwarzschild shadow disk) and the black circle ($a=1$ and $r_s=5.189<3\sqrt{3}$). } 
            \label{k0}
 \end{figure}       

Hence, our results show that, at least for the generalized black‑bounce metrics considered, the Schwarzschild black hole exhibits the largest shadow. This is not surprising, because this topic has been explored in the literature for other kinds of static and spherically symmetric spacetimes \cite{Lu:2019zxb}. In fact, the authors of \cite{Lu:2019zxb} postulated a sequence of inequalities for the parameters characterizing the size of the black hole, i.e.,
\begin{equation}
    \frac{3}{2}r_{sc}\leq r_m\leq\frac{r_s}{\sqrt{3}}\leq 3m,
\end{equation}
where $r_{sc}=2m$. Note that the generalized black-bounce metrics considered here do not satisfy the above inequalities since, as displayed in Fig. \ref{photonspheresk0n2}, the radius of the photon sphere, $r_m$, for $k=0$ and $n=2$ metric may be bigger than $3m$. Although the Schwarzschild black hole remains with the largest shadow size.

}
\section{Conclusions}\label{conclusions}

In this work, we have analyzed light deflection and gravitational lensing in $k-n$ generalized black-bounce spacetimes, characterized by two parameters, $k$ and $n$. In the weak-field gravitational regime, we succeeded in demonstrating that the angular deflection within generalized black-bounce metrics coincides with the usual SV metric up to the second order in the parameters, $a$ and $m$. Conversely, in the strong-field regime the critical impact parameter and the bending angle acquire nonlinear dependence on the bounce parameter $a$ for the cases ($k=0, n=2$) and ($k=2, n=1$), leaving light-bending effects more evident as $a$ grows. Using the lens equations for \textit{Sagittarius} A*, we demonstrated how relativistic image observables depart from their Schwarzschild values as $a/m$ increases. Finally, by computing the shadow radii for several black-bounce models, we confirmed that the Schwarzschild black hole produces the largest shadow size. 

As an unfolding of this work, we intended to study the light deflection for spinning generalized black-bounce metrics and the rotation effects on their shadows. We plan to carry out these investigations in a forthcoming paper.

\textbf{Acknowledgments.} 
This work was partially supported by Conselho Nacional de Desenvolvimento Cient\'{i}fico e Tecnol\'{o}gico (CNPq). The work by A. Yu. P. has been partially supported by the CNPq project No. 303777/2023-0. The work by P. J. P.  has been partially supported by the CNPq project No. 307628/2022-1.

\bibliographystyle{ieeetr}
\bibliography{bibliografia.bib}
\end{document}